\documentclass[twocolumn,tighten]{aastex62}
\hypersetup{colorlinks,%
 citecolor=blue,%
 linkcolor=blue}
\usepackage[normalem]{ulem}
\usepackage{commath}

\definecolor{fgreen}{rgb}{1,0.6,0.1}

\newcommand{\kms}{km\,s$^{-1}$\,}
\newcommand{\mcg}{MCG+07-20-052\,}

\received{\today}
\revised{\today}
\accepted{\today}
\submitjournal{AJ}

\shorttitle{Dwarf-Dwarf merger}
\shortauthors{Paudel et al.}

\begin{document}
\title{MCG+07-20-052: Interacting dwarf pair in a group environment}

\author[0000-0003-2922-6866]{Sanjaya Paudel}
\affil{Department of Astronomy and Center for Galaxy Evolution Research, Yonsei University, Seoul 03722, Korea}
\email{sjyoon0691@yonsei.ac.kr (SJY)\\
}
\author{Chandreyee Sengupta}
\affiliation{Purple Mountain Observatory, Chinese Academy of Sciences, Nanjing, 210034, China}
\author[0000-0002-1842-4325]{ Suk-Jin Yoon}
\affiliation{Department of Astronomy and  Center for Galaxy Evolution Research, Yonsei University, Seoul 03722, Korea}
\author{ Daya Nidhi Chhatkuli}
\affiliation{Central Department of Physics, Tribhuvan University, Kirtipur, Kathmandu, Nepal}

\begin{abstract}

We present an observational study of the interacting pair of dwarf galaxies, \mcg, in the vicinity of Milky Way mass spiral galaxy NGC\,2998. \mcg is located at a sky-projected distance of 105 kpc from NGC\,2998 and the two have a relative line-of-sight velocity of 60 \kms. We observed tidal tail-like extensions on both members (D1 and D2) of the interacting pair \mcg. The interacting dwarf galaxies, D1 and D2, have B-band absolute magnitudes of $-$17.17 and $-$17.14 mag, respectively, and D2 is significantly bluer than D1. We obtained HI 21 cm line data of the NGC\,2998 system using the Giant Metrewave Radio Telescope (GMRT) to get a more detailed view of the neutral hydrogen (HI) emission in the interacting dwarf galaxies and in the galaxy members of the NGC\,2998 group.  Evidence of a merger between the
dwarf galaxies in the MCG+07-20-052 is also present in the HI kinematics and morphology where we find that the HI is mostly concentrated around D2, which also shows a higher level of star-forming activity and bluer $g-r$ color index compared to  D1.  In addition, we detect extended tenuous HI emission around another member galaxy, NGC\,3006, located close to the \mcg-pair at a sky-projected distance of 41 kpc.   We compare here our results from the \mcg pair-NGC\,2998 system with other known LMC-SMC-Milky Way type systems and discuss the possible origin of the dwarf-dwarf interaction.

\end{abstract}


\section{Introduction}

The theory of large-scale structure formation with $\Lambda$ cold dark matter ($\Lambda$CDM) cosmology predicts that a major mass assembly galaxies happens in a hierarchical way and in this hierarchy low-mass galaxies play a crucial role. They are, indeed, the dominant population at all redshifts and simulations predict that dwarf galaxies experience on average three major mergers in their lifetime \citep{Fakhouri10}. While dwarf-dwarf mergers were expected to be more common in early universe \citep{Klimentowski10,Fitts18}, recent observations have shown that they are also present in current epoch. The interaction between dwarf galaxies in isolated environments are frequently reported, however they are rare around massive host \citep{Stierwalt15,Paudel18b,Kado-Fong19}.

\begin{figure*}
\centering
\includegraphics[width=17cm]{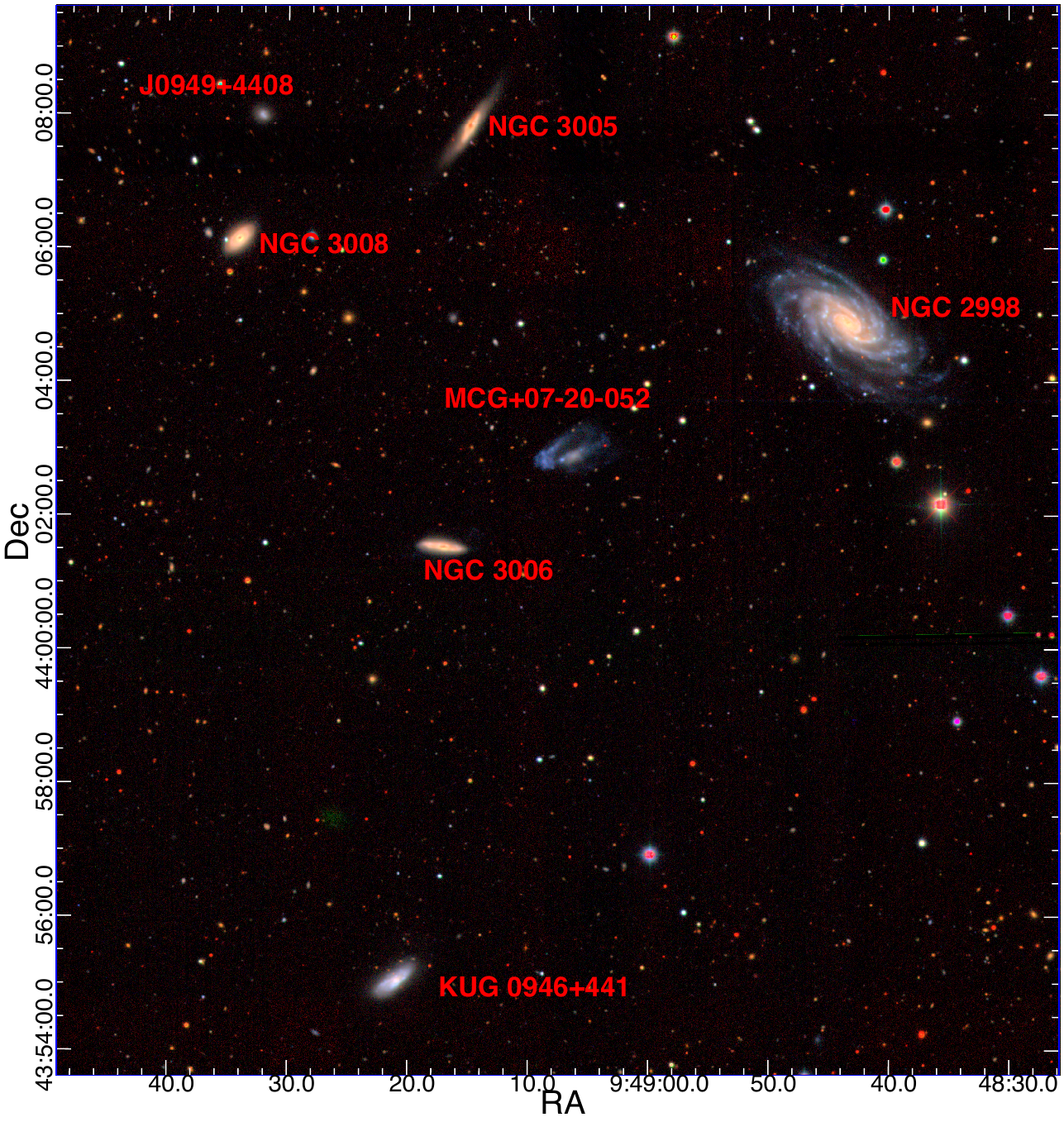}
\caption{Optical view NGC\,2998 seen from the SDSS $g-r-z$ combined tri-color image. The field of view is 14\arcmin$\times$15\arcmin. The centre is adjusted to include all galaxies in the field of view and they are marked.  }
\label{ngc}
\end{figure*}

Massive galaxy mergers have been studied in great detail in both observation and numerical simulations \citep{Barnes92,Naab09,Duc11}. The dwarf-dwarf interactions, on the other hand, are only starting to be explored as a population \citep{Stierwalt15,Pearson16,Besla18,Paudel18b}. This is partly because such systems are possibly not common in the local universe or they are fainter, which makes them harder to detect. The presence of tidal tails, shells, and streams have been studied in great detail around massive galaxies, which provides an unequivocal proof of merger origin of these galaxies \citep{Duc15}. However for dwarf galaxies, such observations are few.  The small shell-like feature at the Fornax dwarf galaxy detected by \cite{Coleman04} indicated merger and recently \cite{Paudel16} presented a detail analysis of shell features early-type dwarf galaxies in the Virgo cluster suggesting merger origin.

Gas-rich merger of massive galaxies leading to the bursts of star formation plays an important role in the stellar mass growth and morphological evolution \citep{Mihos94,Wilman13}. However, the physical processes that trigger enhanced star-formation activity in the low-mass galaxies are still not well understood. Often to explain starburst activity in Blue Compact Dwarf galaxies (BCDs), similar processes as  massive galaxies, eg. the mergers and interactions, have been proposed \citep{Noeske01,Bekki08,Lee09,Privon17}. For the  Large and Small Magellanic clouds (LMC and SMC), interaction is a frequently mentioned reason in the literature to explain the enhanced  star formation activity  \citep{Harris09,Glatt10}. The LMC-SMC system is located in the group environment in vicinity of the Milky Way (MW).  A number of studies have shown that such interaction in group environment is not common and the pair can be quickly disrupted by the host tidal potential \citep{Robotham12,Deason14}. In this paper, we present yet another dwarf-dwarf interacting system ( \mcg), located in the vicinity of NGC\,2998.

\begin{table*}
\centering
\begin{tabular}{cccccccccc}
\hline
Galaxy & RA & Dec & z & M$_{B}$ & $g-r$  &D& SFR & M$_{*}$ & M$_{HI}$\\
 & deg & deg & & mag & mag & kpc& log(M$_{\sun}$/yr) & log(M$_{\sun}$) & log(M$_{\sun}$)\\
 \hline
NGC 2998	   & 147.1825   & 44.0815  & 0.01595  & $-$20.94 & 0.43 &   000 & 	 0.39    & 10.58 & 10.67 \\
NGC 3006	   & 147.3221   & 44.0259  & 0.01600  & $-$18.28 & 0.66 &   152 & 	 $-$1.03 & 9.91  & 9.43 \\
NGC 3005	   & 147.3125	& 44.1313 & 0.01515   & $-$17.99 & 0.75 &   141 &    $-$1.49 & 9.95  & \\  
NGC 3008	   & 147.3927	& 44.1026 & 0.01579   & $-$18.65 & 0.75 &   214 &    $-$1.66 & 10.21 & \\   
KUG 0946+441   & 147.3383	& 43.9179 & 0.01585   & $-$18.68 & 0.33 &   228 &    $-$0.86 & 9.51  & \\  
J0949+4408	   & 147.3849	& 44.1338 & 0.01512   & $-$16.18 & 0.37 &   212 &    $-$1.29 & 8.58  & \\  
D1	           & 147.2771   & 44.0483 & 0.01575  & $-$17.17 & 0.23 &   101 & 	 $-$0.78 & 8.73  & \\
D2	       	   & 147.2870	& 44.0470 &  		  & $-$17.14 & 0.05 &   111 & 	 $-$0.66 & 8.42  & \\
\mcg		   & 		    & 		  &  		  & $-$17.94 & 0.18 &   105 & 	 $-$0.40 & 8.96  & 9.18\\

\hline
\end{tabular}
\caption{Physical properties galaxies in the NGC\,2998 system. We list galaxies name and their position in sky (RA \& Dec) in column 1-3, respectively. The last row, \mcg represents a combine value of D1 and D2. The listed redshifts, in the 4 column,  are obtained from the SDSS spectroscopy. The B-band magnitudes, in  the 5 column, is converted from $g-$band magnitudes and corrected for foreground Galactic extinction. A typical error on magnitudes are 0.01 mag in all optical band photometry. The star-formation rates are derived from FUV flux measured in the GALEX all-sky survey images. The stellar mass, listed in column 8, is derived from the SDSS $r-$band magnitudes in which we expect a systematic uncertainty of 0.2 dex, see the text. 
 }
\label{phtab}
\end{table*}

\section{Interacting dwarf}
As our primary interest is to find low-mass interacting galaxies in the nearby Universe, we carried out a systematic search for such objects in the local volume (z $<$ 0.02). We published the most extended catalog of interacting dwarf galaxies \citep{Paudel18a}, which comprises 177 interacting dwarf galaxies. These galaxies are selected by visual inspection of color images of the two wide-field optical surveys (SDSS-III and the Legacy Survey).  We select them according to their observed low-surface-brightness features that are likely the result of an interaction between dwarf galaxies. The parent sample of dwarf galaxies is selected from NED with a magnitude cut M$_{r}$ $>$ $-$19 mag.  We found that a significant majority of interacting dwarfs are located in the isolated environment and less then 10 percent of them are located in the in vicinity of giant, MW mass or more massive, galaxies. Recently, \cite{Kado-Fong19} also published a sample of interacting dwarf galaxies where they study the star formation and host properties. They also found that dwarf galaxies that host tidal debris are systematically blue, indicating merger induced star formation, which agree with \cite{Paudel18a} findings where an overwhelming majority of interacting dwarf galaxies are star-forming and blue.

 To maximize the number of systems where the interacting galaxies are in spatial and kinematic proximity similar to the LMC-SMC-MW, we defined an LMC-SMC-MW analog as the interacting dwarf located within a 300 kpc sky-projected distance from the MW mass galaxy with a relative line-of-sight velocity less than 300 \kms. Indeed this simple criteria does not entirely reflect the actual LMC-SMC orbital and spatial properties. LMC and SMC are located relatively close, $<$60 kpc, to the MW compared to our 300 kpc  projected distance of interacting dwarf from the their host. One of the main motivations for using this criteria is that we do not have three dimensional information of our system whereas in the case of LMC and SMC, we know their detailed spatial and orbital information in six dimensions.

Among the LMC-SMC-MW analogs found using our criteria, we selected a few systems to study further in detail with HI 21 cm observations. \mcg, is the second object in this series. Previously, we studied UGC\,4703 interacting pair near the isolated spiral NGC\,2718 \citep{Paudel17}, where we discussed the similarity of UGC\,4703 pair galaxies to the LMC-SMC system.

\subsection{\mcg}
MCG+07-20 is located at the sky position RA = 09:49:06.46, Dec. = +44:02:54.35 and a redshift z = 0.01575. It is an interacting pair of star-forming dwarf galaxies located in a group environment. The most massive galaxy in the group, NGC\,2998, is a MW mass spiral. NGC\,2998 group is a relatively dense group which have at least 7 member galaxies of stellar mass larger than 10$^{8}$ M$_{\sun}$ within the 300 kpc sky-projected radius and relative line-of-sight radial velocity $\pm$300 \kms from the central galaxy, NGC\,2998. \mcg is the nearest neighbour to the spiral galaxy
NGC2998 at a sky-projected distance of 105 kpc \footnote{We use NED quoted distance of 60 Mpc for the NGC 2998 group. The distance is derived from the Tully-Fisher relation \citep{Tully16}.} and they have a relative line-of-sight radial velocity of 60 \kms.

\begin{figure}
\centering
\includegraphics[width=8.5cm]{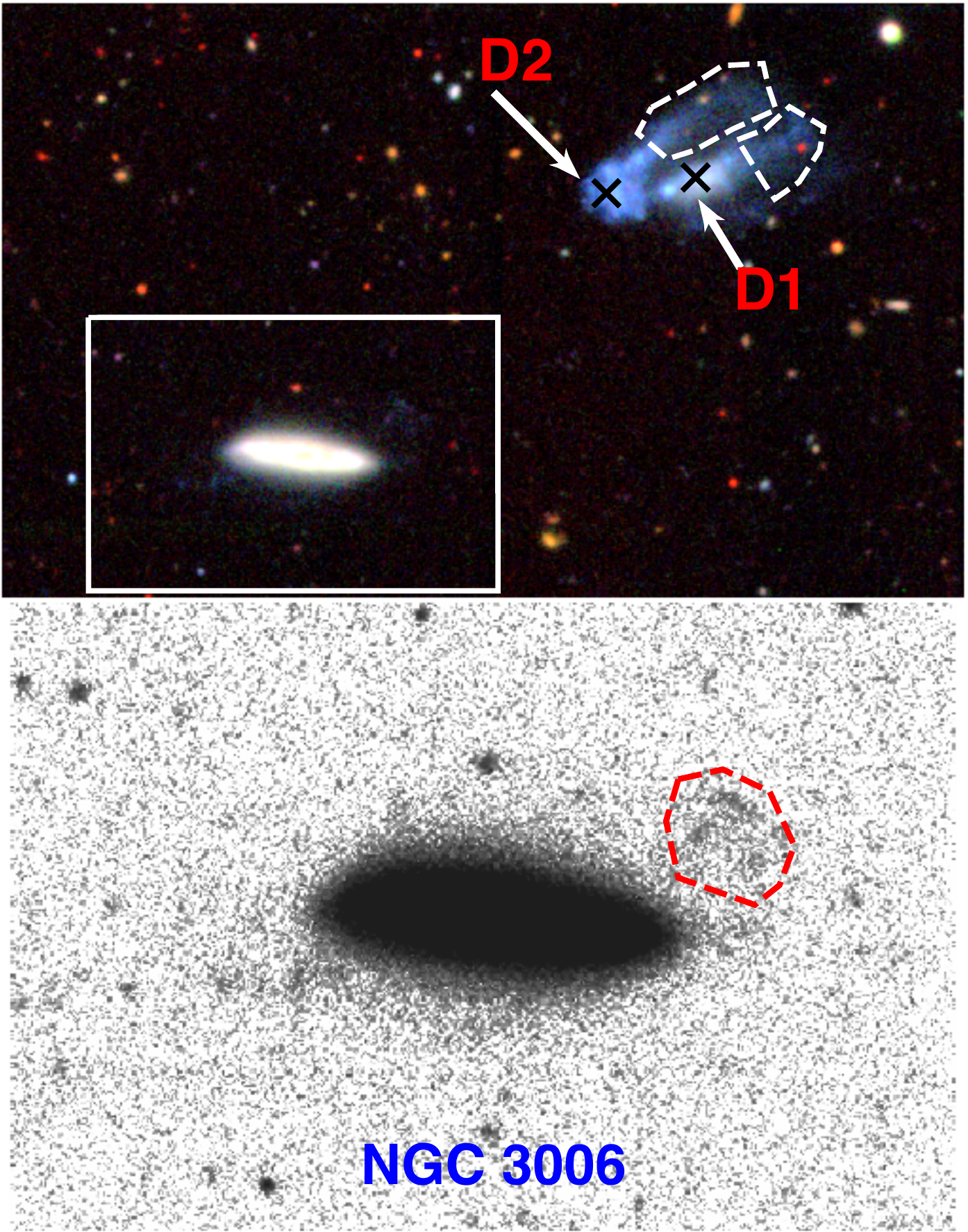}
\caption{A zoom-in view of  the region around \mcg and NGC\,3006. The upper panel image is reproduced same way as Figure \ref{ngc} which have a field view of  4.5\arcmin$\times$3\arcmin. Here we marked the two interacting dwarf galaxies (D1 \& D2) in \mcg. The supposed centre, calculated as centroid, of D1 and D2 are also marked by black crosses and their separation is 25.77\arcsec. In lower panel image, we show further zoom-in of NGC\,3006 and this image is reproduced by co-adding legacy g-r-z band images to gain signal at the low-surface regions. The polygons, white in top and red in lower panel, represent the regions of stellar stream we have chosen for the photometric measurement, see the text.} 
\label{ddm}
\end{figure}

\section{Data analysis}
\subsection{Analysis of archival data}
Optical band photometric measurements were done in the archival images available at the SDSS-III Data Archive Server (DAS)  \citep{Abazajian09}. Although,  deeper optical images are also available from the Legacy survey \citep{Blum16}, we did not use them for the photometric measurement because their flux calibration and sky-background subtraction are not as good as the SDSS images. In this paper, we used the Legacy survey images only for display purpose. The SDSS images were already bias-subtracted, flatfield-corrected, sky-subtracted, and flux-calibrated. We used a tool provided by the SDSS-III to create mosaic image with a large field view and no further effort has been made on the image reduction part. However, we used a simple and similar approach to subtract the sky-background count as in our previous publications \citep{Paudel18a,Paudel18b}. 

In Figure \ref{ngc}, we show the Legacy $g-r-z$ combined tri-color image with a field view of 14\arcmin$\times$15\arcmin of an area around NGC\,2998 which covers all member galaxies of the group. \mcg is located in between the NGC\,3006 and NGC\,2998 and  two tidal tail like extensions can be seen  emanating from the interacting dwarf pair along the direction of NGC\,2998. The sky-projected separation between NGC\,3006 and \mcg  is 41 kpc and relative line-of-sight velocity between the two is 100 \kms. The relative line-of-sight velocities of D1 and D2 from  NGC\,2998 are 18 and 59 \kms, respectively.

In Figure \ref{ddm}, we show a zoom-in region around \mcg and NGC\,3006. We named the two interacting dwarf galaxies in \mcg pair as D1 and D2 located to the east and west respectively. Their sky-projected separation is 7.5 kpc which we measured calculating the centroid of each individual, see cross marks at Figure \ref{ddm}. Although both galaxies look like a typical blue star-forming galaxy, the observed color distribution shows that  the main body of D1 is slightly redder and smooth. On the other hand, the main body of D2 is clumpy and its morphology mimics a typical tadpole galaxy with prominently blue star-forming head and a low-surface brightness tail \citep{Elmegreen12}. A careful look at NGC\,3006 reveals a low-surface brightness (slightly above the detection limit) extension towards \mcg which is significantly blue compared to the main body of the galaxy. The region has been marked with broken line in Figure 2.

To measure the photometric parameters of each individual galaxies, we performed aperture photometry in the SDSS $g$ and $r-$ band filter images. We chose different aperture sizes for each individual galaxies depending on their extension, but kept them fixed for different filters. The sizes of apertures were selected visually and they were wide enough to secure all the flux in the region of interest. Before the doing aperture photometry, we masked all unrelated foreground and background objects manually. Foreground Galactic extinction correction was done applying an extinction map of \cite{Schlafly11} and no correction has been made for internal extinction and cosmological deeming, the later being actually insignificant.

The result of photometric measurements are listed in Table \ref{phtab}.  We converted  the SDSS  $g$-band magnitudes to B-band magnitudes using an equation $B = g +0.227\times(g-r) - 0.337$ provided by the SDSS color transformation tool\footnote{http://www.sdss3.org/dr8/algorithms/sdssUBVRITransform.php}. To derive the stellar mass, we used the $r-$band magnitude with an appropriate mass-to-light ratio (M/L). The M/L is obtained using an equation $log(M/L) =  -0.306 +1.097\times(g-r)$.  This assumes a simple stellar population with a single burst of star-formation to derive a relation between mass-to-light ratio and galaxy color. Indeed, galaxy star-formation histories are complex.  However, it is shown that scatter in the mass-to-light ratio derived from different star-formation history for a given color is $\sim$0.2 dex, see \S4.3 \cite{Zhang17}. Therefore, we consider $\pm$0.2 dex as a typical conservative error on our stellar mass estimates.

The interacting dwarfs, D1 and D2, have a similar brightness with B-band the absolute magnitudes $-$17.17 and $-$17.14 mag, respectively, but D2 is significantly bluer. Their $g-r$ color indexes are 0.23 and 0.05 mag, respectively.  Overall, a combined (D1+D2 a.k.a \mcg) photometry, with a large aperture which covers the both, yields a B-band absolute magnitude $-$17.94 mag. NGC\,3006 has a B-band absolute magnitude $-$18.28 mag and it turns out much redder than interacting dwarf pair with a $g-r$ color index of 0.66 mag. The estimated stellar mass of D1, D2 and NGC\,3006 are 5.37 $\times$10$^{8}$, 2.63 $\times$10$^{8}$ and 8.12 $\times$10$^{9}$ M$_{\sun}$, respectively. That gives a stellar mass ratio between the interacting dwarfs pair is $\sim$2:1.

We measured the $g-r$ color index of the tidal tail like extensions of D1 and D2, marked with broken line in Figure \ref{ddm}. We found the tidal tails of both the dwarfs to have similar $g-r$ color index of 0.22 mag. And for NGC\,3006, we found the low surface brightness eastern extension, also marked with broken line in Figure \ref{ddm}, to be significantly bluer compared to its main body, with a $g-r $ color index of 0.14 mag.

\begin{figure*}
\centering
\includegraphics[width=17cm]{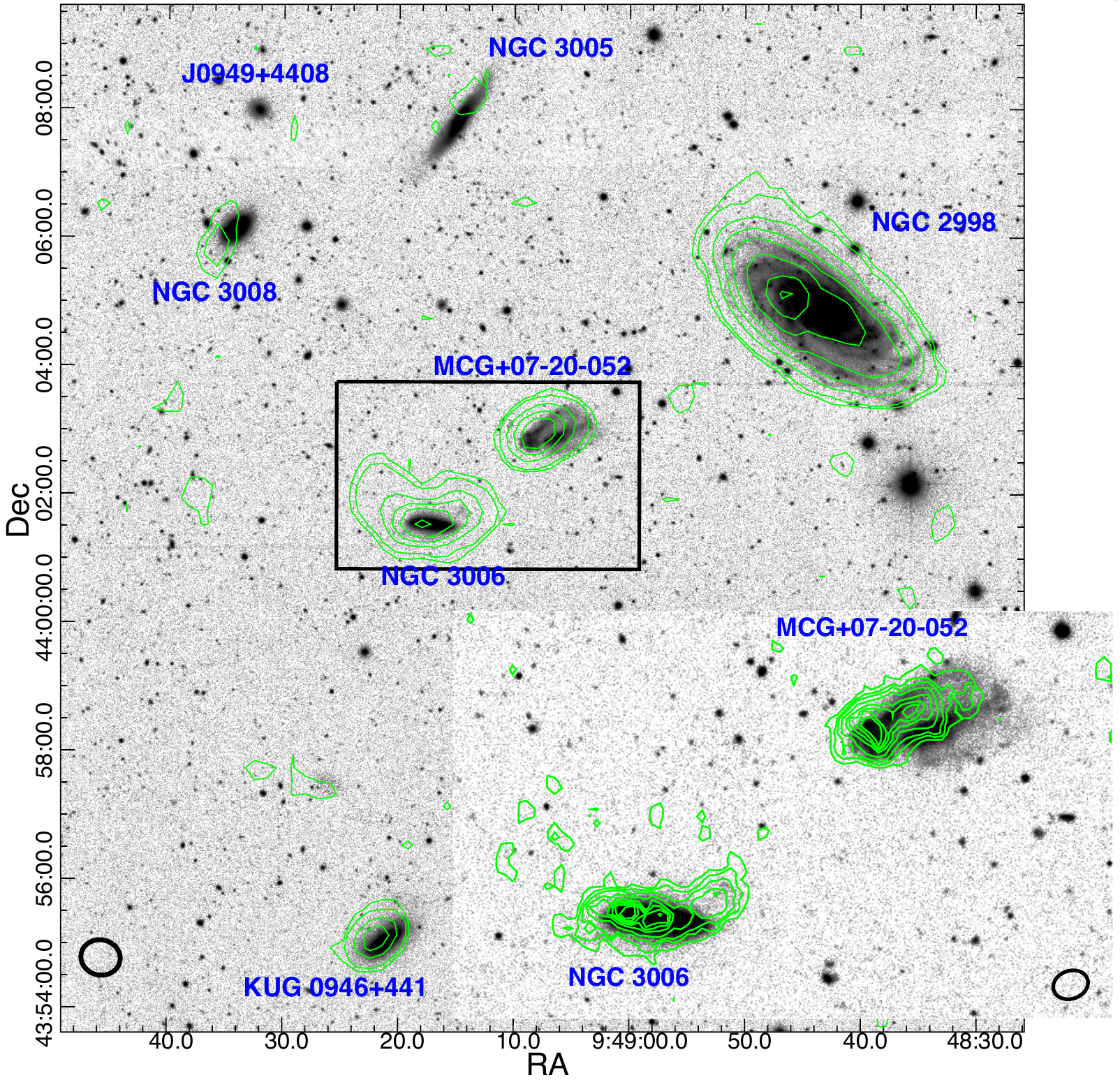}
\caption{Integrated HI contours from the GMRT low-resolution map overlaid on the Legacy co-added g-r-z band images. The field of view of the image is similar to the Figure \ref{ngc}. The HI column density levels are N(H I) = 10$^{20}$ $\times$(0.4, 0.9, 2.7, 4.5, 6.3, 9.0, 11.7, 13.5) cm$^{-2}$. A black ellipse at the bottom left corner represents beam size of GMRT observation. In the inset we show a zoom-in view of a region highlighted by black box with a field of view similar to Figure \ref{ddm}. The overlaid  HI contour is obtained from high-resolution map which has a beam size of 15.3\arcsec$\times$11.9\arcsec, see bottom right corner. The HI column density levels are N(H I) = 10$^{20}$ $\times $(2.5, 4.3, 6.1, 9.0, 12.0, 13.9, 15.1, 16.3, 18.1) cm$^{-2}$.}
\label{himap}
\end{figure*}

To calculate ongoing star-formation rates we used the Galaxy Evolutionary Explorer (GALEX) images. The area around NGC\,2998 are covered by the GALEX all-sky survey \citep{Martin05} and we downloaded intensity map from the archive. Although, the all-sky survey images are not deep exposure, but we still find a reasonable detection in FUV band images for all galaxies in the NGC\,2998 system. We performed aperture photometry as done in the SDSS images in both FUV images. We derived the star-formation rates from the FUV flux using a calibration provided by \cite{Kennicutt98}. The values are listed in Table \ref{phtab}.

\begin{figure}
\centering
\includegraphics[width=8.5cm]{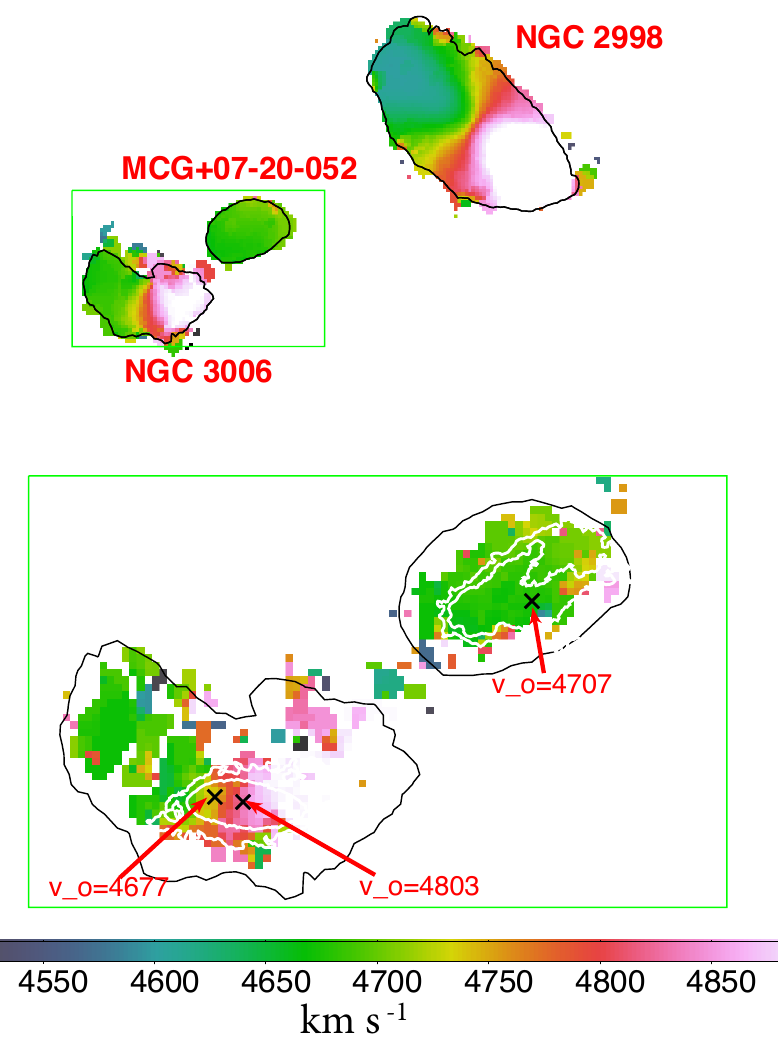}
\caption{ Top: The HI velocity field from the GMRT low-resolution cube.  The black contour traces the lowest column density contour of the low resolution integrated HI image (N(HI)=0.4$\times$10$^{20}$ cm$^{-2}$). 
Bottom:  We show a zoom in view of the region highlighted by a green box. This shows the high resolution velocity field. The white contours traces the optical g-band light at the level  surface brightness 25 and 26.5 mag arsec$^{-2}$ for the inner and outer, respectively. The black contour traces the lowest column density contour of the low resolution integrated HI image, same as the top panel. The black crosses represent the SDSS fiber spectroscopy targeted position, where optical velocities are available.   }
\label{vmap}
\end{figure}
The SDSS fibre spectroscopy has targeted the brighter galaxy, D1, of interacting pair and its measured optical redshift is z=0.0156. We obtain the optical spectrum from the the SDSS archives.  The SDSS 3\arcsec diameter fibre is placed at a relatively red part of central body and the spectrum continuum is not high signal to noise ratio. However, the emission from early Balmer lines ( H$\alpha$, H$\beta$  H$\gamma$ and H$\delta$) are clear. H$\alpha$ equivalent width measured in the SDSS fibre spectrum is 15 \AA{} and this value is not high enough to consider the galaxy as a star-burst, this is typical for local Blue Compact Dwarf galaxies \citep[BCDs][]{Meyer14}. Emission line metallicity derived from ration of H$\alpha$/[NII] is 8.2  \citep{Marino13} and star-formation rate derived from H$\alpha$ emission line flux is 0.001 M$_{\sun}$/yr which is significantly lower than SFR derived from FUV flux. However we note that  3\arcsec diameter fibre spectroscopy only represent the central part of the galaxy and SFR derived from FUV flux is summed over the entire galaxy.

\begin{table}
\caption{GMRT observation details. }
\begin{tabular}{ll}
\hline
Frequency                                 &             1398.3949 MHz \\
Observation date                          &             30 Jan - 1 Feb, 2019   \\     
Phase calibrator (flux density)           &             0834+555 \\                       
Observation time                          &             15.0 h        \\
Primary beam                              &             24\arcsec at 1420 MHz    \\
Low-resolution beam                       &            37.0\arcsec $\times$ 32.9\arcsec (PA = -12.3$^{\circ}$)  \\   
  
High-resolution beam                      &             15.3\arcsec $\times$ 11.9\arcsec (PA = 20.3$^{\circ}$) \\           
rms for low-resolution map                &             1.1 mJy beam$^{-1}$\\    
rms for high-resolution map               &             0.9 mJy beam$^{-1}$   \\                 
RA (pointing centre)                      &             09h 49m 07.6s \\           
Dec (pointing centre)                     &             44d 02m 52.0s \\        
\hline
\end{tabular}
\label{gmrtobs}
\end{table}

\subsection{HI 21-cm observations}\label{radio}

HI observations of \mcg were carried out using the Giant Metrewave Radio Telescope (GMRT\footnote{http://www.ncra.tifr.res.in/ncra/gmrt}) located at Pune, India.  The system was observed between January 30 -- February  1, 2019 as a part of our observing proposal `GMRT HI mapping of LMC-SMC-MW analog'. A 16 MHz bandwidth was used yielding a velocity resolution $\sim$7 \kms. The GMRT primary beam  at L band is 24$^{\prime}$ and the synthesised beams  of the images presented in the paper are 37.0$^{\prime\prime}$ $\times$ 32.9$^{\prime\prime}$ (low resolution) and 15.3$^{\prime\prime}$ $\times$ 11.9$^{\prime\prime}$ (high resolution).  At the adopted distance of \mcg, 60 Mpc, 37\arcsec and 15\arcsec sample 11 kpc and 4 kpc respectively. The data was analysed using the software {\tt AIPS}\footnote{http://www.aips.nrao.edu}. After flagging bad data for radio frequency interference (RFI) or from malfunctioning antennas, the uv data was calibrated and continuum subtracted. Image cubes of various resolutions were then made using the {\tt AIPS} task {\tt IMAGR}.  Finally the integrated HI, velocity field and velocity dispersion maps were made from the HI cubes using the {\tt AIPS} task {\tt MOMNT}.  Further details of the observations are given in Table \ref{gmrtobs}.

Figure \ref{himap} shows a low resolution integrated HI map overlaid on the SDSS $g-r-z$ coadded image. Significant HI  emission is detected in four galaxies NGC\,2998, \mcg, NGC\,3006 and KUG 0946+441.  NGC\,3005 and NGC\,3008 show  marginal detections.  The HI masses of the interacting dwarf pair \mcg and its two significant neighbours NGC\,3006 and NGC\,2998,  were estimated from our observations and are listed in Table 1.  Lack of spatial resolution limited us from estimating individual masses of D1 and D2. The HI disk in NGC\,2998 seem quite regular and unperturbed while both \mcg and NGC\,3006 show extended and irregular HI morphologies. The high resolution (15.3\arcsec $\times$ 11.9\arcsec) image of a part of the field, with NGC\,3006 and \mcg is presented in the inbox. The outermost contour here is $\sim$ 2.5$\times$10$^{20}$ cm$^{-2}$ compared to the outermost contour of 0.4$\times$10$^{20}$ cm$^{-2}$ tracing the diffuse gas in the low resolution image. In \mcg, the high resolution  image reveals that the HI peaks in the merger system coincide mostly with D2. While it was not possible to resolve emission from individual galaxies, two HI peaks were detected. The stronger peak coincides with the head of the "tadpole" D2, and the other one is seen in the region between D1 and D2, see the inset of Figure \ref{himap}. 
HI distribution in NGC\,3006 is extended significantly beyond the optical size of the galaxy. The HI disk of NGC\,3006 is almost U-shaped and  extends towards \mcg, though no HI bridge was detected between the NGC\,3006 and \mcg.

\begin{figure}
\centering
\includegraphics[width=8.5cm]{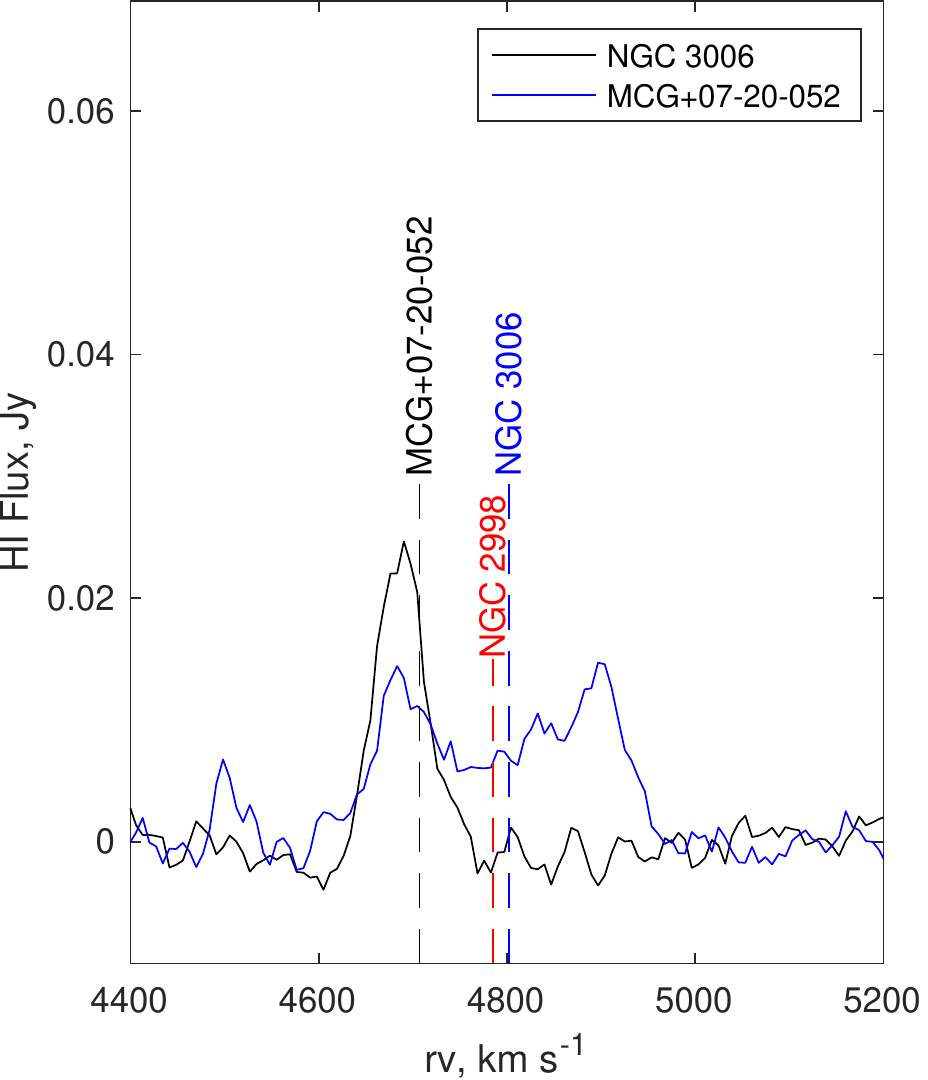}
\caption{ Integrated H I spectrum of \mcg and NGC\,3006. To show the optical velocities, we draw a vertical dash line for each galaxies.}
\label{hispec}
\end{figure}

Figure  \ref{vmap} shows the HI velocity field of NGC\,2998, the \mcg pair and NGC\,3006. In top panel, we show the velocity field obtained from the low-resolution HI cube. In bottom panel, we show a zoom in view of the region highlighted by a green box in the top panel and it shows the high-resolution velocity field. As seen in the low resolution image, NGC\,2998 does not seem to be kinematically affected by any of the other galaxies. HI in the \mcg seem to be merged with bulk of the gas shifted towards D2. We see no signs of rotation or clear velocity gradient in D1 and D2 and they are indistinguishable even in the high resolution image.
The velocity field of NGC\,3006 seems to be preserving its rotation while showing an extended U-shaped morphology, an asymmetry possibly caused by some past interactions. The gas kinematic centre seems to be coincident with the optical centre. The HI spectrum of NGC\,3006 show a regular rotating feature as we can see  the classic double-horn profile and it's centre perfectly matches with the optical velocity of the galaxy centre, Figure \ref{hispec} blue line. SDSS spectroscopy exists for two pointings in NGC\,3006 -one at the centre and another to its west, which have been marked by the black crosses. We find that the gradient in optical velocity of NGC\,3006  matches with that of the HI velocity, suggesting that the gas and stellar masses share same sense of rotation. However, we note that the spatial resolution of our observations, specially the HI data, is not sufficient to make strong claims about the matching gradient, nonetheless the matching spectral line width combined with  the velocity gradient information from the optical and HI maps, suggest an alignment of gas and stellar kinematics in NGC\,3006.

This probably indicates that NGC\, 3006 has not undergone a strong interaction with any of its neighbours. Being in the group,  interactions with its neighbours may have affected the outer edges of its HI disk, but they were never strong enough to disrupt its gas disk or disturb its rotation pattern. Additionally, while the HI in NGC\,3006 is asymmetric and somewhat shows an extension towards \mcg, in the velocity space they seem to be separated by $\sim$100 \kms, suggesting they might have suffered a  fast  and weak interaction in recent past.

\begin{table}
\begin{tabular}{cccccc}
\hline
Galaxy & M$_{B}$ & D & M$_{*}$ & SFR & HI$_{mass}$\\
 & mag & kpc & M$_{\sun}$ &M$_{\sun}$/yr & M$_{\sun}$\\
\hline
\multicolumn{6}{c}{MW}\\
\hline
MW  & -21.17 & 0 & 5$\times$10$^{10}$ &  0.68 - 1.45 & 8$\times$10$^{9}$\\
LMC & -18.60 & 50 & 2.3$\times$10$^{9}$ & 0.2 & 4.4$\times$10$^{8}$\\
SMC & -17.20 & 60 & 5.3$\times$10$^{8}$ & 0.04 & 4$\times$10$^{8}$\\
\hline
\multicolumn{6}{c}{NGC 2718}\\
\hline
NGC 2718 & -21.01 & 0 &  7$\times$10$^{10}$ &0.97 & 1.12$\times$10$^{10}$\\
UGC 4703  & -18.0 & 81 & 1.8 $\times$10$^{9}$ & 0.2 & 1.4 $\times$ 10$^{9*}$\\
UGC 4703B & -16.5 & 104 & 5.4$\times$10$^{8}$ & 0.03 &\\
\hline
\multicolumn{6}{c}{NGC\,2998}\\
\hline
NGC 2998 & -20.94 & 0 &  3.8$\times$10$^{10}$ &2.45 & 8.7$\times$10$^{9}$\\
D1  & -17.17 & 105 & 5.3 $\times$10$^{8}$ & 0.16 & 1.5 $\times$ 10$^{9*}$\\
D2 & -17.14 & 112 & 2.6$\times$10$^{8}$ & 0.21 &\\
NGC\,3006 & -18.28 &  157 &  8.1$\times$10$^{9}$ & 0.09 & 2.6 $\times$ 10$^{9}$\\
\hline
\end{tabular}
\caption{Comparison with LMC-SMC, UGC470 and \mcg. HI masses of NGC\,2998 system have been estimated from the GMRT spectra. The table is reproduced from our previous study on UGC470 in NGC\,2718 group  \citep{Paudel18a}.\\
$^{*}$ Combined H I mass of interacting pair.}
\label{lmccomb}
\end{table}

\section{Discussion}

\subsection{Comparison with the LMC-SMC-MW system and Interaction}
The ongoing interaction between  LMC and SMC, indeed, is happening in a group environment in the vicinity of Milky-Way, although a satellite pair of  LMC-SMC mass around MW mass host is not commonly seen \citep{Robotham12}. Cosmological volume numerical simulations predict that there is less than 10 per cent chance that MW mass halo would host two subhalos of mass of Magellanic clouds \citep{Boylan-Kolchin10,Tollerud11}. Similarly, analyzing the Galaxy And Mass Assembly (GAMA) catalog of galaxies, \cite{Robotham12} estimated the probability of such systems are less than 5\%.  Following these statistics, the merger probability of LMC-SMC morphology dwarf galaxy satellites around the MW mass host maybe even smaller. While analyzing the statistics, we have found $\approx$10 percent (19 out of 177) of interacting dwarf are located within the 700 kpc sky-projected radius from the MW mass halo and among them only  four are qualified as LMC-SMC-MW analogs which are located within the 300 kpc radius. 
Different authors have used different criteria to search LMC-SMC analog.  \cite{Robotham12} selected pair of dwarf galaxies located within the spatial and velocity radius of 100 kpc  $\pm$400 \kms, respectively, from the host and \cite{Liu11}  selected pair of dwarf galaxies located within the spatial and velocity radius of 100 kpc  $\pm$300 \kms, respectively, from the host. Whereas, we selected the interacting dwarfs, not simply pair, located within the 300 kpc sky-projected radius and $\pm$300 \kms line-of slight radial velocity from the host. Given different selection criteria for LMC-SMC analog in the literatures and this study, we refrain from making any strong claims and only present a qualitative comparison. 

\begin{figure}
\centering
\includegraphics[width=8.5 cm]{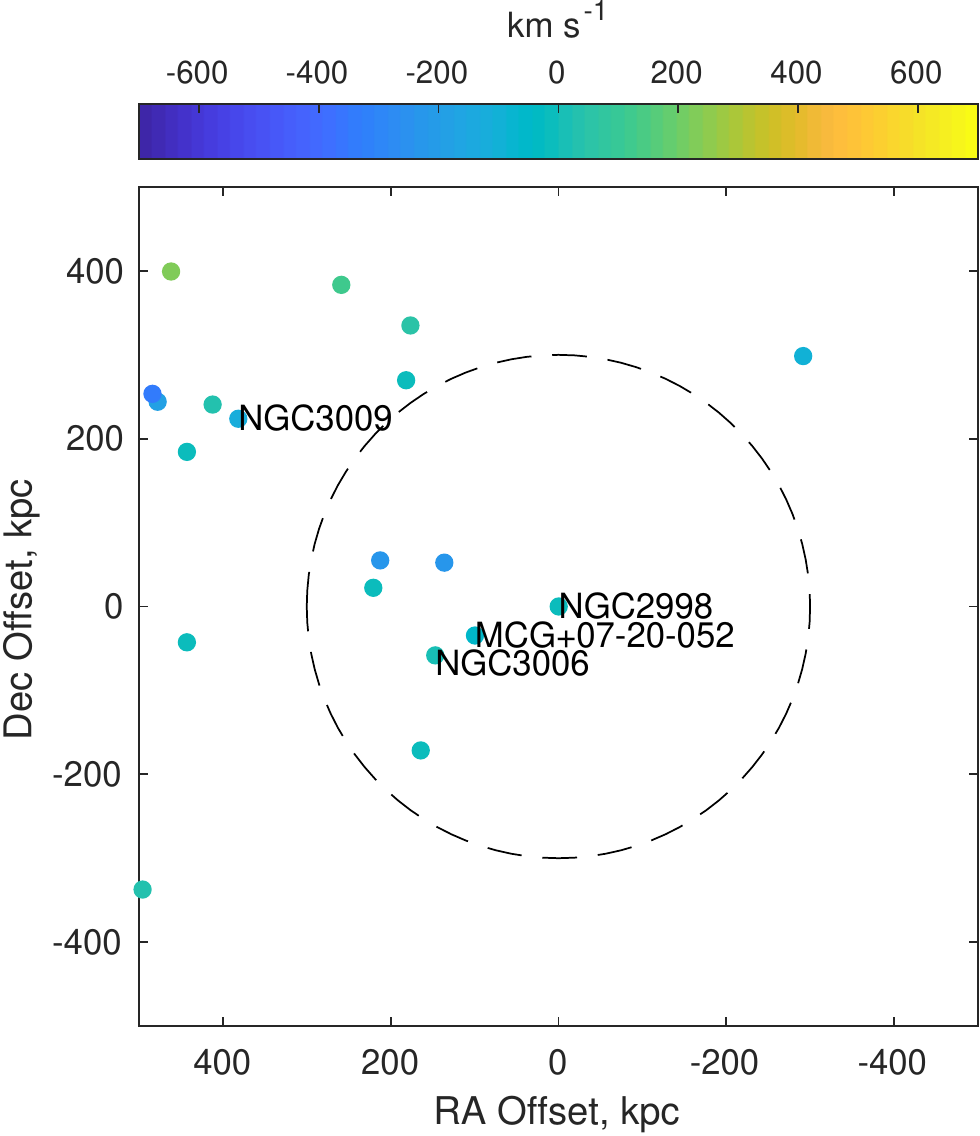}
\caption{Distribution of galaxies around NGC\,2998. Where we sampled all galaxies available at the NED within the velocity range of $\pm$700 \kms with respect to the line of light velocity of NGC\,2998. The x-axis and y-axis are offsets in RA and DEC from NGC\,2998, respectively, and both have a unit of kpc. The symbol color represents their relative line-of-sight velocities with respect to NGC\,2998, according to the color-bar at the top of the plot. We highlight immediate, 300 kpc sky-projected radius region around the NGC\,2998 by a dotted red circle.  }
\label{alsky}
\end{figure}

\mcg  is located at  a sky projected distance of 105 kpc from the NGC\,2998 which have the stellar mass similar to the MW. We list a direct comparison of physical parameters between the LMC-SMC-MW system and the NGC\,2998-\mcg in Table \ref{lmccomb}.  In comparison to MW and NGC 2718, NGC\,2998 is located in a relatively dense environment where we find several bright (brighter than SMC) galaxies around it. There are at least seven galaxies with confirmed redshift information within a sky projected radius of 300 kpc around NGC\,2998. In addition to the galaxies presented in Figure 1, we find that NGC\,3009 is located at a sky projected distance of 442 kpc which is a magnitude fainter than NGC\,2998.

In Figure \ref{alsky}, we show the immediate environment of NGC\,2998 within an area of $\pm$500 kpc in sky. The position and radial velocities of galaxies are obtained from NED. We can see many galaxies within our selected sky area and this certainly make the immediate environment of NGC\,3009 significantly different from MW and NGC 2718.  There are only two bright star-forming galaxies located around the MW (LMC and SMC) and NGC\,2718 (UGC\,4703 and UGC\,4703B). This, indeed, shows that  NGC\,2998 is located in a significantly denser environment than MW and NGC\,2718.

In NGC\,2718, the interacting satellites were connected through a stellar stream where we identified a blue star-forming region, a possible candidiate Tidal Dwarf galaxy (TDG) and their sky-projected separation was 20 kpc. However in \mcg we do not find such stellar stream connecting the interacting dwarfs D1 and D2, and also their separation is quite small, almost overlapping each other. Lack of spatial resolution prevented us from estimating the gas loss in terms of individual HI deficiencies  for D1 or D2, but we do notice a hint of an HI shift towards D2. So we tried to estimate the average gas content of the MCG pair, although we acknowledge the uncertainties are large in such an estimate. Assuming the net HI content of 1.6$\times$10$^9$ M$_{\odot}$ is shared equally by D1 and D2, and the average optical diameter of the galaxies is 0.5\arcmin (8.8 kpc), the HI surface density (M$_{HI}$/{D$_l$}$^{2}$) value turns out to be $\sim$7.0. Comparing this with the average HI surface density value of an unperturbed field galaxy of similar morphological type and taking into account the uncertainty on that number \citep{Haynes84}, we conclude the MCG pair shows no significant HI deficiency. Compared to the  reported cases of gas transfer to the outskirts in forms of extended tails in dwarf-dwarf pairs \cite{Pearson16}, the MCG pair shows no major gas loss and seems to retain majority of its gas close to its optical extent. The two dwarfs D1 and D2 seem to be  interacting leading to gas transfer or a projected shift of the bulk of the gas from D1 to D2, but there seems to be no massive gas loss from the pair into the IGM.

Our HI data is fairly deep, with the lowest HI contour tracing $\sim$ 4$\times$10$^{19}$ cm$^{-2}$. But unlike the Magellanic Stream (MS) we do not detect any HI extension/ emission connecting  NGC\,2998 and the \mcg. Instead, we find that HI distribution around NGC\,3006, another LMC mass dwarf galaxy, to be asymmetric and stretched towards  \mcg. The warped but unperturbed rotation pattern of NGC\,3006, as observed in the HI as well as optical data, suggests it may be undergoing or may have undergone a flyby interaction with its neighbour galaxy \mcg. We derive tidal index, $\Theta$\footnote{$\Theta = log(\frac{M_{*}[\times10^{11}]}{D^{3}{proj}})$ where M$_{*}$ is the stellar mass of host and D$_{proj}$ is the projected distance from the dwarf \citep{Karachentsev13}.}, on the \mcg pair from the  NGC\,2998 and NGC\,3006, i.e $\Theta_{NGC\,2998} \sim 2.4$ and $\Theta_{NGC\,3006} \sim 3.07$. Which, indeed, shows that the tidal influence of NGC\,3006 is significantly larger than that of NGC\,2998. The LMC/SMC tidal index is significantly larger than these values, which is  3.7 \citep{Pearson16}.

The head-tail elongation seen in optical in D1 and D2 as well as the pushed `U' shaped HI morphology seen in NGC\,3006 could usually be seen as signs of ongoing Ram Pressure Stripping \citep[RPS][]{Abramson11} in the group. Commonly known as jelly-fish galaxies, ongoing RPS galaxies also show compression of HI on the opposite side of an elongated tail and depending on the strength of the ram pressure, the HI distribution show signs of outer edge gas stripping, eg. JO206 \citep[see][]{Ramatsoku19}. Also RPS mainly affects the gas distribution before it affects the stellar disk. In \mcg, while the tadpole like feature is visible in optical, we find no hint of compression or shrinking of the HI disk and the HI distribution is more extended than the galaxy's stellar distribution in both NGC\,3006 and \mcg(see white contour in Figure \ref{vmap}).  Therefore this extended, merged and asymmetric HI distribution in \mcg suggests that the observed tidal features on both HI gas and stellar population, including the `tadpole' like appearance of the dwarfs may have originated mainly from tidal interaction and not from RPS. Ram pressure alone could not be responsible for the optical features of D1 and D2, rather it could be tidal assisted ram pressure which is more commonly found to influence galaxy evolution in group environments \citep{Davis97,Sengupta07,Besla12,Pearson18}. In contrast, in the LMC the presence of stars well beyond the HI distribution in the leading edge suggests that RPS played a strong role and \cite{Salem15} concludes that LMC's disk was truncated by the peak ram pressure at pericentric passage.


\section{Conclusions}

We present GMRT HI images and optical photometry data of the galaxies in the  NGC\,2998 group. From the data analysis and their discussion in the broader context, we draw the following conclusions.\\
1)  We identify \mcg, a pair of dwarf galaxies of LMC-SMC mass, located in the vicinity of MW mass spiral galaxy NGC\,2998.\\
2)  HI is mostly concentrated around only one member of the interacting pair, i.e D2, which also shows a higher level of star-forming activity and bluer g$-$r color index compared to the other member of interacting pair, i.e D1. \\
3) We detect, extended tenuous emission of HI around another member galaxy, NGC\,3006, located close to the \mcg at a sky-projected distance of 41 kpc.\\
4) We present a comparative analysis between LMC-SMC-MW, NGC 2718-UGC 4703 and NGC\,2998 -\mcg. We found, NGC\,2998 is located in a relatively denser  environment compared to the MW and NGC\,2718.  In addition, unlike the Magellanic Stream we do not detect any HI extension connecting the \mcg and the host NGC\,2998.

\section{Acknowledgements}
We thank the staff of the GMRT who have made these observations
  possible. GMRT is run by the National Centre for Radio Astrophysics
  of the Tata Institute of Fundamental Research
S.P. acknowledges support from the New Researcher Program (Shinjin grant No. 2019R1C1C1009600) through the National Research Foundation of Korea and by the Yonsei University Research Fund (Post-doctoral Researcher Supporting Program) of 2019 (No. 2019-12-0005). 
S.-J.Y. acknowledges support from the Mid-career Researcher Program (No. 2019R1A2C3006242) and the SRC Program (Center for Galaxy Evolution Research; No. 2017R1A5A1070354) through the National Research Foundation of Korea.
This study is based on the archival images and spectra from the Sloan Digital Sky Survey (the full acknowledgment can be found at http://www.sdss.org/ collaboration/credits.html). The Legacy Surveys imaging of the DESI footprint is supported by the Director, Office of Science, Office of High Energy Physics of the U.S. Department of Energy under Contract No. DE-AC02- 05CH1123, by the National Energy Research Scientific Computing Center, a DOE Office of Science User Facility under the same contract; and by the U.S. National Science Foundation, Division of Astronomical Sciences under Con- tract No. AST-0950945 to NOAO. We also made use of the GALEX all-sky survey imaging data. The GALEX is operated for NASA by the California Institute of Technology under NASA contract NAS5-98034.


\begin{thebibliography}{}
\expandafter\ifx\csname natexlab\endcsname\relax\def\natexlab#1{#1}\fi
\providecommand{\url}[1]{\href{#1}{#1}}
\providecommand{\dodoi}[1]{doi:~\href{http://doi.org/#1}{\nolinkurl{#1}}}
\providecommand{\doeprint}[1]{\href{http://ascl.net/#1}{\nolinkurl{http://ascl.net/#1}}}
\providecommand{\doarXiv}[1]{\href{https://arxiv.org/abs/#1}{\nolinkurl{https://arxiv.org/abs/#1}}}

\bibitem[{{Abazajian} {et~al.}(2009){Abazajian}, {Adelman-McCarthy},
  {Ag{\"u}eros}, {Allam}, {Allende Prieto}, {An}, {Anderson}, {Anderson},
  {Annis}, {Bahcall}, \& et~al.}]{Abazajian09}
{Abazajian}, K.~N., {Adelman-McCarthy}, J.~K., {Ag{\"u}eros}, M.~A., {et~al.}
  2009, \apjs, 182, 543, \dodoi{10.1088/0067-0049/182/2/543}

\bibitem[{{Abramson} {et~al.}(2011){Abramson}, {Kenney}, {Crowl}, {Chung}, {van
  Gorkom}, {Vollmer}, \& {Schiminovich}}]{Abramson11}
{Abramson}, A., {Kenney}, J.~D.~P., {Crowl}, H.~H., {et~al.} 2011, \aj, 141,
  164, \dodoi{10.1088/0004-6256/141/5/164}

\bibitem[{{Barnes}(1992)}]{Barnes92}
{Barnes}, J.~E. 1992, \apj, 393, 484, \dodoi{10.1086/171522}

\bibitem[{{Bekki}(2008)}]{Bekki08}
{Bekki}, K. 2008, \mnras, 388, L10, \dodoi{10.1111/j.1745-3933.2008.00489.x}

\bibitem[{{Besla} {et~al.}(2012){Besla}, {Kallivayalil}, {Hernquist}, {van der
  Marel}, {Cox}, \& {Kere{\v s}}}]{Besla12}
{Besla}, G., {Kallivayalil}, N., {Hernquist}, L., {et~al.} 2012, \mnras, 421,
  2109, \dodoi{10.1111/j.1365-2966.2012.20466.x}

\bibitem[{{Besla} {et~al.}(2018){Besla}, {Patton}, {Stierwalt},
  {Rodriguez-Gomez}, {Patel}, {Kallivayalil}, {Johnson}, {Pearson}, {Privon},
  \& {Putman}}]{Besla18}
{Besla}, G., {Patton}, D.~R., {Stierwalt}, S., {et~al.} 2018, \mnras, 480,
  3376, \dodoi{10.1093/mnras/sty2041}

\bibitem[{{Blum} {et~al.}(2016){Blum}, {Burleigh}, {Dey}, {Schlegel},
  {Meisner}, {Levi}, {Myers}, {Lang}, {Moustakas}, {Patej}, {Valdes}, {Kneib},
  {Huanyuan}, {Nord}, {Olsen}, {Delubac}, {Saha}, {James}, {Walker}, \& {DECaLS
  Team}}]{Blum16}
{Blum}, R.~D., {Burleigh}, K., {Dey}, A., {et~al.} 2016, in American
  Astronomical Society Meeting Abstracts, Vol. 228, American Astronomical
  Society Meeting Abstracts \#228, 317.01

\bibitem[{{Boylan-Kolchin} {et~al.}(2010){Boylan-Kolchin}, {Springel}, {White},
  \& {Jenkins}}]{Boylan-Kolchin10}
{Boylan-Kolchin}, M., {Springel}, V., {White}, S.~D.~M., \& {Jenkins}, A. 2010,
  \mnras, 406, 896, \dodoi{10.1111/j.1365-2966.2010.16774.x}

\bibitem[{{Coleman} {et~al.}(2004){Coleman}, {Da Costa}, {Bland-Hawthorn},
  {Mart{\'{\i}}nez-Delgado}, {Freeman}, \& {Malin}}]{Coleman04}
{Coleman}, M., {Da Costa}, G.~S., {Bland-Hawthorn}, J., {et~al.} 2004, \aj,
  127, 832, \dodoi{10.1086/381298}

\bibitem[{{Davis} {et~al.}(1997){Davis}, {Keel}, {Mulchaey}, \&
  {Henning}}]{Davis97}
{Davis}, D.~S., {Keel}, W.~C., {Mulchaey}, J.~S., \& {Henning}, P.~A. 1997,
  \aj, 114, 613, \dodoi{10.1086/118497}

\bibitem[{{Deason} {et~al.}(2014){Deason}, {Wetzel}, \&
  {Garrison-Kimmel}}]{Deason14}
{Deason}, A., {Wetzel}, A., \& {Garrison-Kimmel}, S. 2014, \apj, 794, 115,
  \dodoi{10.1088/0004-637X/794/2/115}

\bibitem[{{Duc} {et~al.}(2011){Duc}, {Cuillandre}, {Serra}, {Michel-Dansac},
  {Ferriere}, {Alatalo}, {Blitz}, {Bois}, {Bournaud}, {Bureau}, {Cappellari},
  {Davies}, {Davis}, {de Zeeuw}, {Emsellem}, {Khochfar}, {Krajnovi{\'c}},
  {Kuntschner}, {Lablanche}, {McDermid}, {Morganti}, {Naab}, {Oosterloo},
  {Sarzi}, {Scott}, {Weijmans}, \& {Young}}]{Duc11}
{Duc}, P.-A., {Cuillandre}, J.-C., {Serra}, P., {et~al.} 2011, \mnras, 417,
  863, \dodoi{10.1111/j.1365-2966.2011.19137.x}

\bibitem[{{Duc} {et~al.}(2015){Duc}, {Cuillandre}, {Karabal}, {Cappellari},
  {Alatalo}, {Blitz}, {Bournaud}, {Bureau}, {Crocker}, {Davies}, {Davis}, {de
  Zeeuw}, {Emsellem}, {Khochfar}, {Krajnovi{\'c}}, {Kuntschner}, {McDermid},
  {Michel-Dansac}, {Morganti}, {Naab}, {Oosterloo}, {Paudel}, {Sarzi}, {Scott},
  {Serra}, {Weijmans}, \& {Young}}]{Duc15}
{Duc}, P.-A., {Cuillandre}, J.-C., {Karabal}, E., {et~al.} 2015, \mnras, 446,
  120, \dodoi{10.1093/mnras/stu2019}

\bibitem[{{Elmegreen} {et~al.}(2012){Elmegreen}, {Elmegreen}, {S{\'a}nchez
  Almeida}, {Mu{\~n}oz-Tu{\~n}{\'o}n}, {Putko}, \& {Dewberry}}]{Elmegreen12}
{Elmegreen}, D.~M., {Elmegreen}, B.~G., {S{\'a}nchez Almeida}, J., {et~al.}
  2012, \apj, 750, 95, \dodoi{10.1088/0004-637X/750/2/95}

\bibitem[{{Fakhouri} {et~al.}(2010){Fakhouri}, {Ma}, \&
  {Boylan-Kolchin}}]{Fakhouri10}
{Fakhouri}, O., {Ma}, C.-P., \& {Boylan-Kolchin}, M. 2010, \mnras, 406, 2267,
  \dodoi{10.1111/j.1365-2966.2010.16859.x}

\bibitem[{{Fitts} {et~al.}(2018){Fitts}, {Boylan-Kolchin}, {Bullock}, {Weisz},
  {El-Badry}, {Wheeler}, {Faucher-Gigu{\`e}re}, {Quataert}, {Hopkins},
  {Kere{\v{s}}}, {Wetzel}, \& {Hayward}}]{Fitts18}
{Fitts}, A., {Boylan-Kolchin}, M., {Bullock}, J.~S., {et~al.} 2018, \mnras,
  479, 319, \dodoi{10.1093/mnras/sty1488}

\bibitem[{{Glatt} {et~al.}(2010){Glatt}, {Grebel}, \& {Koch}}]{Glatt10}
{Glatt}, K., {Grebel}, E.~K., \& {Koch}, A. 2010, \aap, 517, A50,
  \dodoi{10.1051/0004-6361/201014187}

\bibitem[{{Harris} \& {Zaritsky}(2009)}]{Harris09}
{Harris}, J., \& {Zaritsky}, D. 2009, \aj, 138, 1243,
  \dodoi{10.1088/0004-6256/138/5/1243}

\bibitem[{{Haynes} \& {Giovanelli}(1984)}]{Haynes84}
{Haynes}, M.~P., \& {Giovanelli}, R. 1984, \aj, 89, 758, \dodoi{10.1086/113573}

\bibitem[{{Kado-Fong} {et~al.}(2019){Kado-Fong}, {Greene}, {Greco}, {Beaton},
  {Goulding}, {Johnson}, \& {Komiyama}}]{Kado-Fong19}
{Kado-Fong}, E., {Greene}, J.~E., {Greco}, J.~P., {et~al.} 2019, arXiv
  e-prints, arXiv:1911.08497.
\newblock \doarXiv{1911.08497}

\bibitem[{{Karachentsev} {et~al.}(2013){Karachentsev}, {Makarov}, \&
  {Kaisina}}]{Karachentsev13}
{Karachentsev}, I.~D., {Makarov}, D.~I., \& {Kaisina}, E.~I. 2013, \aj, 145,
  101, \dodoi{10.1088/0004-6256/145/4/101}

\bibitem[{{Kennicutt}(1998)}]{Kennicutt98}
{Kennicutt}, Jr., R.~C. 1998, \araa, 36, 189,
  \dodoi{10.1146/annurev.astro.36.1.189}

\bibitem[{{Klimentowski} {et~al.}(2010){Klimentowski}, {{\L}okas}, {Knebe},
  {Gottl{\"o}ber}, {Martinez-Vaquero}, {Yepes}, \& {Hoffman}}]{Klimentowski10}
{Klimentowski}, J., {{\L}okas}, E.~L., {Knebe}, A., {et~al.} 2010, \mnras, 402,
  1899, \dodoi{10.1111/j.1365-2966.2009.16024.x}

\bibitem[{{Lee} {et~al.}(2009){Lee}, {Kennicutt}, {Funes}, {Sakai}, \&
  {Akiyama}}]{Lee09}
{Lee}, J.~C., {Kennicutt}, Jr., R.~C., {Funes}, S.~J.~J.~G., {Sakai}, S., \&
  {Akiyama}, S. 2009, \apj, 692, 1305, \dodoi{10.1088/0004-637X/692/2/1305}

\bibitem[{{Liu} {et~al.}(2011){Liu}, {Gerke}, {Wechsler}, {Behroozi}, \&
  {Busha}}]{Liu11}
{Liu}, L., {Gerke}, B.~F., {Wechsler}, R.~H., {Behroozi}, P.~S., \& {Busha},
  M.~T. 2011, \apj, 733, 62, \dodoi{10.1088/0004-637X/733/1/62}

\bibitem[{{Marino} {et~al.}(2013){Marino}, {Rosales-Ortega}, {S{\'a}nchez},
  {Gil de Paz}, {V{\'{\i}}lchez}, {Miralles-Caballero}, {Kehrig},
  {P{\'e}rez-Montero}, {Stanishev}, {Iglesias-P{\'a}ramo}, {D{\'{\i}}az},
  {Castillo-Morales}, {Kennicutt}, {L{\'o}pez-S{\'a}nchez}, {Galbany},
  {Garc{\'{\i}}a-Benito}, {Mast}, {Mendez-Abreu}, {Monreal-Ibero}, {Husemann},
  {Walcher}, {Garc{\'{\i}}a-Lorenzo}, {Masegosa}, {Del Olmo Orozco},
  {Mour{\~a}o}, {Ziegler}, {Moll{\'a}}, {Papaderos},
  {S{\'a}nchez-Bl{\'a}zquez}, {Gonz{\'a}lez Delgado}, {Falc{\'o}n-Barroso},
  {Roth}, {van de Ven}, \& {Califa Team}}]{Marino13}
{Marino}, R.~A., {Rosales-Ortega}, F.~F., {S{\'a}nchez}, S.~F., {et~al.} 2013,
  \aap, 559, A114, \dodoi{10.1051/0004-6361/201321956}

\bibitem[{{Martin} {et~al.}(2005){Martin}, {Fanson}, {Schiminovich},
  {Morrissey}, {Friedman}, {Barlow}, {Conrow}, {Grange}, {Jelinsky},
  {Milliard}, {Siegmund}, {Bianchi}, \& {Byun}}]{Martin05}
{Martin}, D.~C., {Fanson}, J., {Schiminovich}, D., {et~al.} 2005, \apjl, 619,
  L1, \dodoi{10.1086/426387}

\bibitem[{{Meyer} {et~al.}(2014){Meyer}, {Lisker}, {Janz}, \&
  {Papaderos}}]{Meyer14}
{Meyer}, H.~T., {Lisker}, T., {Janz}, J., \& {Papaderos}, P. 2014, \aap, 562,
  A49, \dodoi{10.1051/0004-6361/201220700}

\bibitem[{{Mihos} \& {Hernquist}(1994)}]{Mihos94}
{Mihos}, J.~C., \& {Hernquist}, L. 1994, \apjl, 425, L13,
  \dodoi{10.1086/187299}

\bibitem[{{Naab} \& {Ostriker}(2009)}]{Naab09}
{Naab}, T., \& {Ostriker}, J.~P. 2009, \apj, 690, 1452,
  \dodoi{10.1088/0004-637X/690/2/1452}

\bibitem[{{Noeske} {et~al.}(2001){Noeske}, {Iglesias-P{\'a}ramo},
  {V{\'{\i}}lchez}, {Papaderos}, \& {Fricke}}]{Noeske01}
{Noeske}, K.~G., {Iglesias-P{\'a}ramo}, J., {V{\'{\i}}lchez}, J.~M.,
  {Papaderos}, P., \& {Fricke}, K.~J. 2001, \aap, 371, 806,
  \dodoi{10.1051/0004-6361:20010446}

\bibitem[{{Paudel} \& {Sengupta}(2017)}]{Paudel17}
{Paudel}, S., \& {Sengupta}, C. 2017, \apjl, 849, L28,
  \dodoi{10.3847/2041-8213/aa95bf}

\bibitem[{{Paudel} {et~al.}(2018{\natexlab{a}}){Paudel}, {Sengupta}, \&
  {Yoon}}]{Paudel18b}
{Paudel}, S., {Sengupta}, C., \& {Yoon}, S.-J. 2018{\natexlab{a}}, \aj, 156,
  166, \dodoi{10.3847/1538-3881/aadb8d}

\bibitem[{{Paudel} {et~al.}(2018{\natexlab{b}}){Paudel}, {Smith}, {Yoon},
  {Calder{\'o}n-Castillo}, \& {Duc}}]{Paudel18a}
{Paudel}, S., {Smith}, R., {Yoon}, S.~J., {Calder{\'o}n-Castillo}, P., \&
  {Duc}, P.-A. 2018{\natexlab{b}}, \apjs, 237, 36,
  \dodoi{10.3847/1538-4365/aad555}

\bibitem[{{Paudel} {et~al.}(2016){Paudel}, {Smith}, {Duc}, {C{\^o}t{\'e}},
  {Cuillandre}, {Ferrarese}, {Blakeslee}, {Boselli}, {Cantiello}, {Gwyn},
  {Guhathakurta}, {Mei}, {Mihos}, {Peng}, {Powalka}, {S{\'a}nchez-Janssen},
  {Toloba}, \& {Zhang}}]{Paudel16}
{Paudel}, S., {Smith}, R., {Duc}, P.-A., {et~al.} 2016, ArXiv e-prints.
\newblock \doarXiv{1611.03561}

\bibitem[{{Pearson} {et~al.}(2016){Pearson}, {Besla}, {Putman}, {Lutz},
  {Fern{\'a}ndez}, {Stierwalt}, {Patton}, {Kim}, {Kallivayalil}, {Johnson}, \&
  {Sung}}]{Pearson16}
{Pearson}, S., {Besla}, G., {Putman}, M.~E., {et~al.} 2016, \mnras, 459, 1827,
  \dodoi{10.1093/mnras/stw757}

\bibitem[{{Pearson} {et~al.}(2018){Pearson}, {Privon}, {Besla}, {Putman},
  {Mart{\'\i}nez-Delgado}, {Johnston}, {Gabany}, {Patton}, \&
  {Kallivayalil}}]{Pearson18}
{Pearson}, S., {Privon}, G.~C., {Besla}, G., {et~al.} 2018, \mnras, 480, 3069,
  \dodoi{10.1093/mnras/sty2052}

\bibitem[{{Privon} {et~al.}(2017){Privon}, {Stierwalt}, {Patton}, {Besla},
  {Pearson}, {Putman}, {Johnson}, {Kallivayalil}, {Liss}, \&
  {Titans}}]{Privon17}
{Privon}, G.~C., {Stierwalt}, S., {Patton}, D.~R., {et~al.} 2017, \apj, 846,
  74, \dodoi{10.3847/1538-4357/aa8560}

\bibitem[{{Ramatsoku} {et~al.}(2019){Ramatsoku}, {Serra}, {Poggianti},
  {Moretti}, {Gullieuszik}, {Bettoni}, {Deb}, {Fritz}, {van Gorkom},
  {Jaff{\'e}}, {Tonnesen}, {Verheijen}, {Vulcani}, {Hugo}, {J{\'o}zsa},
  {Maccagni}, {Makhathini}, {Ramaila}, {Smirnov}, \& {Thorat}}]{Ramatsoku19}
{Ramatsoku}, M., {Serra}, P., {Poggianti}, B.~M., {et~al.} 2019, \mnras, 487,
  4580, \dodoi{10.1093/mnras/stz1609}

\bibitem[{{Robotham} {et~al.}(2012){Robotham}, {Baldry}, {Bland-Hawthorn},
  {Driver}, {Loveday}, {Norberg}, {Bauer}, {Bekki}, {Brough}, {Brown},
  {Graham}, {Hopkins}, {Phillipps}, {Power}, {Sansom}, \&
  {Staveley-Smith}}]{Robotham12}
{Robotham}, A.~S.~G., {Baldry}, I.~K., {Bland-Hawthorn}, J., {et~al.} 2012,
  \mnras, 424, 1448, \dodoi{10.1111/j.1365-2966.2012.21332.x}

\bibitem[{{Salem} {et~al.}(2015){Salem}, {Besla}, {Bryan}, {Putman}, {van der
  Marel}, \& {Tonnesen}}]{Salem15}
{Salem}, M., {Besla}, G., {Bryan}, G., {et~al.} 2015, \apj, 815, 77,
  \dodoi{10.1088/0004-637X/815/1/77}

\bibitem[{{Schlafly} \& {Finkbeiner}(2011)}]{Schlafly11}
{Schlafly}, E.~F., \& {Finkbeiner}, D.~P. 2011, \apj, 737, 103,
  \dodoi{10.1088/0004-637X/737/2/103}

\bibitem[{{Sengupta} {et~al.}(2007){Sengupta}, {Balasubramanyam}, \&
  {Dwarakanath}}]{Sengupta07}
{Sengupta}, C., {Balasubramanyam}, R., \& {Dwarakanath}, K.~S. 2007, \mnras,
  378, 137, \dodoi{10.1111/j.1365-2966.2007.11748.x}

\bibitem[{{Stierwalt} {et~al.}(2015){Stierwalt}, {Besla}, {Patton}, {Johnson},
  {Kallivayalil}, {Putman}, {Privon}, \& {Ross}}]{Stierwalt15}
{Stierwalt}, S., {Besla}, G., {Patton}, D., {et~al.} 2015, \apj, 805, 2,
  \dodoi{10.1088/0004-637X/805/1/2}

\bibitem[{{Tollerud} {et~al.}(2011){Tollerud}, {Boylan-Kolchin}, {Barton},
  {Bullock}, \& {Trinh}}]{Tollerud11}
{Tollerud}, E.~J., {Boylan-Kolchin}, M., {Barton}, E.~J., {Bullock}, J.~S., \&
  {Trinh}, C.~Q. 2011, \apj, 738, 102, \dodoi{10.1088/0004-637X/738/1/102}

\bibitem[{{Tully} {et~al.}(2016){Tully}, {Courtois}, \& {Sorce}}]{Tully16}
{Tully}, R.~B., {Courtois}, H.~M., \& {Sorce}, J.~G. 2016, \aj, 152, 50,
  \dodoi{10.3847/0004-6256/152/2/50}

\bibitem[{{Wilman} {et~al.}(2013){Wilman}, {Fontanot}, {De Lucia}, {Erwin}, \&
  {Monaco}}]{Wilman13}
{Wilman}, D.~J., {Fontanot}, F., {De Lucia}, G., {Erwin}, P., \& {Monaco}, P.
  2013, \mnras, 433, 2986, \dodoi{10.1093/mnras/stt941}

\bibitem[{{Zhang} {et~al.}(2017){Zhang}, {Puzia}, \& {Weisz}}]{Zhang17}
{Zhang}, H.-X., {Puzia}, T.~H., \& {Weisz}, D.~R. 2017, \apjs, 233, 13,
  \dodoi{10.3847/1538-4365/aa937b}

\end{thebibliography}

\end{document}